\documentstyle[12pt]{article}

\begin{document}
\title{Dynamically generated heavy scalar fields in the Standard Model of Electroweak Interactions}
\author{Bing An Li\\
Department of Physics and Astronomy, University of Kentucky\\
Lexington, KY 40506, USA}

\maketitle
\begin{abstract}
It is shown that the gauge fixings of the Z and the W fields and three scalars
are via nonconserved axial-vector and charged vector currents of massive fermions
dynamically generated by fermion masses in the SM of EW interactions. 
The top quark mass provides enough strength for this chiral symmetry breaking.
The masses of the three scalars are determined to be about $10^{14}$ GeV. 
These scalars have negative probability. They are the nonperturbative solutions of the SM.
A new perturbation theory with dynamically generated and fixed gauge fixings is constructed. 
The Faddeev-Popov procedure is not invoked.
\end{abstract}
\newpage

In the SM of electroweak interactions(EW)[1] there are spontaneous symmetry breaking, Higgs, and Faddeev-Popov procedure. 
The theory is successful in many aspects. Other theoretical approaches:
Supersymmetry models, Technichcolor, little Higgs etc., have been proposed to modify the SM.
In Ref.[2] the effect of top quark condensate has been studied.
A new mechanism of chiral symmetry breaking has been found in Ref.[3]:
the masses and the gauge fixings of the Z and the W fields and three scalars are 
via nonconserved axial-vector and charged vector currents 
of fermions dynamically generated by fermion masses in EW theory.
The mass of the top quark provides enough strength for chiral symmetry breaking. 
In this paper this new mechanism is applied to the SM of EW theory.
It is shown that the scalars are the nonperturbative solutions of the SM. 
A brief review of previous study of the new mechanism is presented first.

This mechanism is inspired by Weinberg's second sum rule[4], \(m^2_a=2m^2_\rho\).
$a_1$ and $\rho$ mesons are chiral partners and they are degenerate under chiral symmetry.
Obviously, Weinberg's second sum rule implicates a new mechanism of chiral symmetry breaking.
This new mechanism has been studied by
a chiral field theory of pseudoscalar, vector,
and axial-vector mesons[5]. Based on current algebra and QCD the Lagrangian of this theory
has been constructed as 
\begin{equation}
{\cal L}=\bar{\psi}(x)(i\gamma\cdot\partial+\gamma\cdot v
+\gamma\cdot a\gamma_{5}-mu(x))\psi(x)
+{1\over 2}m^{2}_{0}(\rho^{\mu}_{i}\rho_{\mu i}+
\omega^{\mu}\omega_{\mu}+a^{\mu}_{i}a_{\mu i}+f^{\mu}f_{\mu}),
\end{equation}
where \(u=e^{(i\pi+i\eta_0)\gamma_5}\), m is the constituent quark mass which originates in quark condensate.
The Lagrangian(1) has global $SU(2)_L\times SU(2)_R$ symmetry and the $a_1$ and the $\rho$ fields
have the same mass in Eq.(1).
Integrating out the quark fields, the Lagrangian of the mesons is
derived, in which $a_1$ and $\rho$ are nonabelian gauge fields. This theory is phenomenologically successful.
The current quark masses can be added into Eq.(1). The masses of the pseudoscalar mesons are obtained from explicit chiral 
symmetry breaking by the current quark masses. 
The masses of the vector mesons originate in dynamical chiral symmetry breaking and \(m^2_\rho=6m^2\) is obtained. 
The mass relation between the $a_1$ and the $\rho$ mesons is derived 
\begin{equation}
(1-{1\over 2\pi^2 g^2})m^2_a=2m^2_\rho,
\end{equation}
where g is a universal coupling constant. Comparing with Weinberg's second
sum rule, there is an additional factor. In obtaining the Weinberg's second sum rule an assumption has been made in Ref.[4].
This new formula fits the data well. Therefore, besides the explicit and the dynamical chiral symmetry breaking 
of QCD a new one(2) emerges from this theory(1). 
In Eq.(1) the $a_1$ field is coupled to the axial-vector current of the massive fermion. 
Because this current is not conserved[6] the vacuum polarization of the $a_1$ field is expressed as
\begin{equation}
\Pi_{\mu\nu}^{ij}(q^2)=\delta_{ij}\{F_1(q^2)(q_\mu q_\nu-q^2 g_{\mu\nu})+F_2(q^2)q_\mu q_\nu+{1\over2}\Delta m^2 g_{\mu\nu}\}.
\end{equation}
In Ref.[5] $\Pi_{\mu\nu}$ has been calculated to the fourth order in covariant derivatives.
$F_2$ and $\Delta m^2$ are the results of the nonconserved axial-vector current of the massive quark.
The Weinberg's second sum rule is obtained from $\Delta m^2=m^2_\rho$.
The factor $1-{1\over 2\pi^2 g^2}$ comes from $F_1$ which is different from the corresponding function 
of the $\rho$ field(1). This difference originates in the constituent quark mass m[5]. 
Besides Eq.(2), the gauge fixing of the $a_1$ field is revealed from $F_2$.
This is a {\bf new mechanism of chiral symmetry breaking}: nonconserved axial-vector current of massive quark 
leads to both the mass difference of $a_1$ and $\rho$ and the gauge fixing of the $a_1$ field.  
It is very important to notice that in the Lagrangian(1) there is no kinetic term for the $a_1$ field and the  
mass of the $a_1$ field is not physical. The physical L of the $a_1$ field is obtained by integrating out the quark fields.
This procedure is equivalent to treat quark loop(3) nonperturbatively. 
The new mechanism has nonperturbative nature.

Detailed study and the consequences of this new chiral symmetry breaking induced by fermion mass are presented in Ref.[3a].  
A model of massless axial-vector field and massive fermion 
has been constructed[3a]. The $a_\mu$-field is coupled to the axial-vector 
current of the massive fermion. There is no gauge invariance and the free Lagrangian of $a_\mu$ field 
cannot be defined at the tree level. The free Lagrangian of the fermion is defined as usual. 
The vacuum polarization of $a_\mu$ field is calculated.
Because the axial-vector current of the massive fermion is not conserved the amplitude of the vacuum polarization 
takes the form of Eq.(3). Both $F_2$ and $\Delta m^2$ are proportional to the fermion mass.
The mass of the $a_\mu$-field and the gauge fixing are dynamically generated from $\Delta m^2$ and $F_2$ respectively. 
The free Lagrangian of the $a_\mu$-field is constructed as  
\[{\cal L}=-{1\over4}(\partial_\mu a_\nu-\partial_\nu a_\mu)^2+\xi(\partial_\mu a^\mu)^2+{1\over2}m^2_a a_\mu^2\]
and it is a Stueckelberg's Lagrangian, where $\xi$ is a constant obtained from $F_2$. 
The $a_\mu$-field has four independent components.
Nonzero $\partial_\mu a^\mu$ leads to a spin-0 field and gauge fixing. 
The new spin-0 field is the nonperturbative solution of this theory.
The mass of the spin-0 field is determined by the finite $F_2$. 
The perturbation theory is constructed. The propagator of a-field with dynamically generated gauge fixing consists 
of two parts: spin-1 and spin-zero. 
There is a minus sign in front of the spin-0 part. Therefore,
the spin-0 field has negative probability. 
Similar mechanism is revealed[3a] from a theory of charged vector fields which are coupled 
to nonconserved charged 
vector currents of massive fermions: the mass and the gauge fixing of the charged vector 
fields and the charged scalar fields are via vacuum polarization dynamically generated from the fermion masses. 
The mass of the charged scalar is explicitly determined and it has negative probability.  
%It is known for a long time that the term, $\frac{q_\mu q_\nu}{M^2}$, in the propagator of a massive vector(axial-vector) field
%has problem at high energies.
%The mechanism of the spontaneous symmetry breaking allows the renormalization gauge to be introduced.
%The study done in Ref.[3] shows that the gauge fixing for the propagator of the massive vector(axial-vector) field can be
%dynamically generated in a theory in which
%the vector(axial-vector) field is either coupled to axial-vector current or charged vector currents of massive fermions.
%In this new mechanism the symmetry is broken by fermion mass and the mass of the gauge field, the gauge fixing, and the scalar
%field are via the new mechanism dynamically generated from the fermion mass. 

The axial-vector and the charged vector currents of the massive fermions exist in the EW theory. 
Therefore, the new mechanism of chiral 
symmetry breaking is already embedded in the theory. On the other hand, there is heavy top quark in the EW theory. 
The mass of the top quark provides enough strength for this chiral symmetry breaking.
%Therefore, there are both the new mechanism induced by the fermion masses and the strength of the symmetry breaking 
%provided by the mass of the top quark in the EW theory.
The masses of the intermediate bosons, the gauge fixings, and three scalars can via the new mechanism[5,3a] 
be dynamically generated without invoking the spontaneous symmetry breaking and the Faddeev-Popov procedure in the EW theory.
 
In Ref.[3a] an EW theory without spontaneous symmetry breaking, Higgs, and Faddeev-Popov procedure has been proposed.
The Lagrangian of the (t,b)-quark generation is constructed as
\begin{eqnarray}
{\cal L}&=&
-{1\over4}A^{i}_{\mu\nu}A^{i\mu\nu}-{1\over4}B_{\mu\nu}B^{\mu\nu}
+\bar{q}_L\{i\gamma\cdot\partial
+{g\over2}\tau_{i}
\gamma\cdot A^{i}+g'{Y\over2}\gamma\cdot B\}
q_{L}\nonumber \\
&&+\bar{q}_{R}\{i\gamma\cdot\partial+g'{Y\over2}\gamma\cdot Bi\}q_{R}
-m_t\bar{t}t-m_b\bar{b}b.
\end{eqnarray}
Obviously, Eq.(4) is part of the Lagrangian of the SM.
The Lagrangian of other generations of fermions can be added too. 
Eq.(4) shows that the gauge fields are massless and gauge symmetries are broken by the fermion masses. Therefore,  
the free Lagrangian of the massless intermediate bosons 
cannot be constructed.
Because of nonconservation of the axial-vector and the charged vector currents of the massive fermions
the fermion loops of the vacuum polarizations of the Z and the W fields are expressed as Eq.(3). The propagators of Z and W
cannot be defined at the tree level, therefore,
the calculation of the vacuum polarizations can only be done at one-fermion loop level[3a].  
Taking the vacuum polarizations of the Z and the W fields into account, the free Lagrangian of the Z and the W
fields is defined[3a] respectively and it is Stueckelberg's Lagrangian.  
After renormalizing the fermion masses, 
\(m^2_W={1\over2}g^2 m^2_t\,\;m^2_Z={1\over2}\bar{g}^2 m^2_t,\;G_F=\frac{1}{2\sqrt{2}m^2_t}\)[3]
are obtained. They agree with data very well. The top quark plays dominant role. The theory predicts
\(\frac{m^2_W}{m^2_Z}=cos^2\theta_W\)[3], which is independent of the scheme of renormalization. 
The functions $F_2's$ are finite and there are no new divergences.
Three scalar fields are dynamically generated from finite $F_2's$, whose masses $\sim10^{14}$GeV are determined.
The propagators of the Z- and the W-fields
with the gauge fixings revealed from $F_2's$ are derived. The intermediate boson loops can only be calculated correctly 
after their propagators are defined by taking the fermion loops into account. 
The intermediate boson loops for the vacuum
polarizations of the intermediate bosons are taken as perturbation.
%\[\Delta^Z_{\mu\nu}=
%\frac{1}{p^2-m^2_Z}\{-g_{\mu\nu}+(1+\frac{1}{2\xi_Z})\frac{p_\mu p_\nu}{
%p^2-m^2_{\phi^0}}\},\;\;\xi_Z=-{m^2_Z\over 2m^2_{\phi^0}}\;\;and\]
%\[\Delta^W_{\mu\nu}=
%\frac{1}{p^2-m^2_W}\{-g_{\mu\nu}+(1+\frac{1}{2\xi_W})\frac{p_\mu p_\nu}{
%p^2-m^2_{\phi_W}}\},\;\;\xi_W=-{m^2_W\over 2m^2_{\phi_W}}.\]
The three scalar fields have negative probability. New perturbation theory with dynamically generated gauge fixings has been
constructed.
The effects of the heavy scalars can be found from the loop diagrams of the physical
processes of the EW theory.
This theory keeps all the success of the SM at the tree level as energies much lower than $10^{14}$ GeV.

The SM is revisited in this paper. The axial-vector and the charged
vector currents of the massive fermions exist in the SM after the spontaneous symmetry breaking and they are not conserved.
Besides the spontaneous symmetry breaking the nonconserved currents of massive fermions lead to another chiral symmetry breaking
which has been already embedded in the SM. The top quark provides enough strength for this chiral symmetry breaking.
According to Ref.[3], when the vacuum polarizations of the Z- and the W-field by the massive fermions are treated nonperturbatively
dynamical generation of the gauge fixings and the three heavy scalar fields is inevitable. 
Therefore, the Feddeev-Popov procedure is no longer required. 
%Why the solutions of scalar fields have not
%been found in the SM before? The key point is that all the EW interactions have been treated perturbatively.
Why the fermion loops of the vacuum polarizations of the Z- and the W-field must be treated nonperturbatively?
In the original SM all the interactions are treated perturbatively and after spontaneous symmetry breaking 
the Z and the W are massive spin-1 fields and each has three degrees of freedoms.
However, after nonperturbative treatment of the fermion loops of the vacuum polarization of the Z- and the W-field each of them 
gains the fourth component-a scalar field, therefore, each has four independent degrees of freedoms.  
A properly constructed perturbation theory shouldn't change the number of independent degrees of freedoms. 
It is known that in QED if all interactions are treated as perturbation, there are no positronium solutions.
Only when the coulomb potential separated from the photon propagator of $ee^+\rightarrow ee^+$ scattering 
is treated nonperturbatively the solutions of positronium can be found. 
The scalar fields are nonperturbative solutions of the SM.
There is another issue. For the vacuum polarizations of the Z- and the W- field
at one-loop level besides the fermion loops there are other loop diagrams: loop diagrams of Higgs and intermediate bosons,
loop diagrams of intermediate bosons only, and Higgs loop. The Higgs loop only contributes to 
the mass term of the intermediate boson. Except the fermion loops all 
other loop diagrams that affect $F_2$ require the propagators of intermediate bosons 
which can only be correctly defined
after the fermion loops of the vacuum polarizations of the Z and the W field are treated nonperturbatively. 
Based on these arguments 
the loop diagrams of fermions are treated nonperturbatively and all others are treated as perturbation. 
%The quark loops with internal lines of Higgs are treated perturbatively too. 

Because of nonconservation of the axial-vector currents of the massive fermions the amplitude of the vacuum 
polarization of the Z-field by the fermions is expressed as
\begin{equation}
\Pi^Z_{\mu\nu}={1\over2}F_{Z1}(q^2)(q_\mu q_\nu-q^2 g_{\mu\nu})+F_{Z2}(q^2)
q_\mu q_\nu+{1\over2}\Delta m^2_Z g_{\mu\nu},
\end{equation}
$F_{Z1}(q^2)$, $F_{Z2}(q^2)$, and $\Delta m^2_Z$ are presented in Ref.[3] at one-fermion loop level. 
Only the axial-vector currents contribute to $F_{Z2}(q^2)$ and $\Delta m^2_Z$.
In the limit $m_f\rightarrow 0$ $F_{Z2}(z)$ and $\Delta m^2_Z$ go to zero. 
$F_{Z1}(z)$ is used to renormalize the Z-field and $\Delta m^2_Z$ is absorbed 
via renormalization by $m^2_Z$. $F_{Z2}(z)$ is finite.  
A constant is separated from $F_{Z2}(z)$
\begin{equation}
F_{Z2}(q^2)=\xi_Z+(q^2-m^2_{\phi^0})G_{Z2}(q^2),
\end{equation}
where $m^2_{\phi^0}$ is the mass of a new neutral spin-0 field
and $G_{Z2}$ is the radiative correction. The new free Lagrangian of the Z-field
is constructed as
\begin{equation}
{\cal L}_{Z0}=-{1\over4}(\partial_\mu Z_\nu-\partial_\nu Z_\mu)^2+\xi_Z(\partial_\mu Z^\mu)^2+{1\over2}m^2_Z Z^2_\mu.
\end{equation}
Eq.(7) is Stueckelberg's Lagrangian. $\partial_\mu Z^\mu$ is an independent spin-0 field
\begin{eqnarray}
\phi^0&=&-{m_Z\over m^2_{\phi^0}}\partial_\mu Z^\mu,\;\;\;
\partial^2\phi^0-{m^2_Z\over2\xi_Z}\phi^0=0,\;\;\; 
Z_\mu=Z'_\mu+{1\over m_Z}\partial_\mu\phi^{0},\;\;\;
\partial_\mu Z'^\mu=0.
\end{eqnarray}
The mass of $\phi^0$ is determined by following equation
\begin{equation}
2m^2_{\phi^0}F_{Z2}|_{q^2=m^2_{\phi^0}}+m^2_Z=0.
\end{equation}
Eq.(9) is dominated by the mass of the top quark and has a solution at very large $q^2$. 
The mass of the $\phi^0$ and the gauge fixing are determined to be
\begin{eqnarray}
m_{\phi^0}&=&m_t e^{\frac{m^2_z}{m^2_t}{16\pi^2\over3\bar{g}^2}+1}
=3.78\times10^{14}GeV,\;\;\;
\xi_Z=-\frac{m^2_Z}{2m^2_{\phi^0}}=-1.18\times10^{-25}.
\end{eqnarray}
%The positive $m_{\phi^0}$ is resulted in negative $\xi_Z$. 
From Eq.(7) the propagator of Z boson is derived
\begin{equation}
\Delta_{\mu\nu}=
\frac{1}{q^2-m^2_Z}\{-g_{\mu\nu}+(1+\frac{1}{2\xi_Z})\frac{q_\mu q_\nu}{
q^2-m^2_{\phi^0}}\}=
\frac{1}{q^2-m^2_Z}\{-g_{\mu\nu}+\frac{q_\mu q_\nu}{
m^2_Z}\}-\frac{1}{m^2_Z}\frac{q_\mu q_\nu}{q^2
-m^2_{\phi^0}}.
\end{equation}
There is a minus sign in front of the propagator of the scalar field(11).
The scalar field has negative probability[3a].

The W boson is coupled to both the charged axial-vector and  
vector currents of the
massive fermions. These currents are not conserved. The vacuum polarization of the W-field by the massive fermions is expressed as
\begin{equation}
\Pi^W_{\mu\nu}=F_{W1}(q^2)(q_\mu q_\nu-q^2 g_{\mu\nu})+2F_{W2}(q^2)
q_\mu q_\nu+\Delta m^2_W
g_{\mu\nu}.
\end{equation}
$F_{W1}$, $F_{W2}$, and $\Delta m^2_W$ are presented in Ref.[3] at one-fermion loop level. $F_{W1}$ is used to renormalize W-field
and the mass of the W boson is renormalized by $\Delta m^2_W$. Both the charged axial-vector and vector currents
contribute to $F_{W2}$ and $\Delta m^2_W$ which are proportional to the masses
of the fermions. 
$F_{W2}$ is finite and a constant is separated from it
\begin{equation}
F_{W2}=\xi_W+(q^2-m^2_{\phi_W})G_{W2}(q^2),
\end{equation}
where $G_{W2}$ is the radiative correction 
and $m^2_{\phi_W}$ is the mass of the charged spin-0 states. 
The free Lagrangian of W-field is defined as
\begin{equation}
{\cal L}_{W0}=-{1\over2}(\partial_\mu W^+\nu-\partial_\nu W^+_\mu)
(\partial_\mu W^-_\nu-\partial_\nu W^-_\mu)+2\xi_W\partial_\mu W^{+^\mu}
\partial_\nu W^{-^\nu}+m^2_W W^+_\mu W^{-\mu}.
\end{equation}
Eq.(14) is Stueckelberg's Lagrangian.
The divergence of the W-field is a new scalar field
\begin{eqnarray}
\phi^{\pm}&=&-\frac{m_W}{m^2_{\phi^W}}\partial_\mu W^{\pm\mu},\;\;
\partial^2(\partial_\mu W^{\pm\mu})-{m^2_W\over2\xi_W}(\partial_\mu
W^{\pm\mu})=0.
\end{eqnarray}
Eqs.(14,15) show that the mass of the charged scalar is the solution of the equation
\begin{equation}
2m^2_{\phi_W}F_{W2}(q^2)|_{q^2=m^2_{\phi_W}}+m^2_W=0.
\end{equation}
Eq.(16) has a solution at very large $q^2$ and top quark mass dominates the solution. The mass of $\phi_W$ and the gauge 
fixing coefficient are determined to be
\begin{eqnarray}
m_{\phi_W}&=&m_t e^{{16\pi^2\over3g^2}{m^2_W\over m^2_t}}
=9.31\times10^{13}GeV,\;\;\;
\xi_W=-{m^2_W\over2m^2_{\phi_W}}=-3.73\times10^{-25}.
\end{eqnarray}
Negative $\xi_W$ leads to positive $m_{\phi_W}$.
The propagator of W-field with dynamically generated gauge fixing is derived from Eq.(14)
\begin{equation}
\Delta^W_{\mu\nu}=
\frac{1}{q^2-m^2_W}\{-g_{\mu\nu}+(1+\frac{1}{2\xi_W})\frac{q_\mu q_\nu}{
q^2-m^2_{\phi_W}}\}=
\frac{1}{q^2-m^2_W}\{-g_{\mu\nu}+\frac{q_\mu q_\nu}{
m^2_W}\}-\frac{1}{m^2_W}\frac{q_\mu q_\nu}{q^2-m^2_{\phi_W}}.
\end{equation}
Once again, there is a minus sign in front of the propagator
of the scalar field. The charged scalar fields have negative probability.

The key functions $F_{2Z, 2W}$ are calculated at one-fermion loop level.
However, arguments for the existence of the scalar field and the gauge fixing 
in a theory in which the current(the gauge field is coupled to) is not conserved can be very general. 
%As a matter of fact, in any theory of EW interactions there are always nonconserved axial-vector
%and charged vector currents of massive fermions. One neutral and two charged scalars are 
%nonperturbative solutions of any theory of EW interactions. 
Because of nonconservation of the current the vacuum polarization of the gauge field should be written as
\begin{equation}
\Pi_{\mu\nu}(q)=f_1(q^2)(q_\mu q_\nu -q^2 g_{\mu\nu})+f_2(q^2)g_{\mu\nu}.
\end{equation}
Defining \(f_2(q^2)-f_2(0)=-F_2(q^2)q^2\), Eq.(19) is rewritten as
\begin{equation}
\Pi_{\mu\nu}(q)=F_1(q^2)(q_\mu q_\nu -q^2 g_{\mu\nu})+F_2(q^2)q_\mu q_\nu+f_2(0)g_{\mu\nu}.
\end{equation}
The function $F_2$ and the constant $f_2(0)$ are the results of the nonconserved current.
$f_2(0)$ is the mass term of the gauge field.
After nonperturbative treatment of the vacuum polarization the gauge fixing and the scalar field are revealed from $F_2$.
Because $F_2$ is unknown the mass
of the scalar field cannot be determined.   

The perturbation theory of the SM needs to be reconstructed. The free Lagrangian of Z and W are shown in Eqs.(7,14).
After $\xi_{Z,W}$ are
separated, the remaining parts of $F_{2Z}$ and $F_{2W}$ are still kept in the perturbation Lagrangian.
Other part of the perturbation theory is the same as in the original perturbation theory of the SM. 
The differences between the new perturbation theory and the original one are:
the gauge parameters of Eqs.(11,18) are dynamically generated and fixed(10,17), and there are no Faddeev-Popov ghosts. 
The Faddeev-Popov procedure used in the original SM is no longer invoked and there is no gauge choosing in this new version 
of the perturbation theory. 
Use of the propagators(11,18)
to calculate the loop diagrams of physical processes is the direct test of the effects of the heavy scalars.
For the tree diagrams at present level of energies the contribution of the scalar fields can be 
ignored. Therefore, the success of the SM at the tree level is kept.
The free Lagrangian of the scalar fields is presented in Ref.[3a]. 
The interactions between the scalar fields and others can be derived.
As an example, the interaction between the charged scalars and the photon is derived
\begin{equation}
{\cal L}={4ie\over m^2_W}F^{\mu\nu}\partial_\mu\phi^+\partial_\nu\phi^-.
\end{equation}
The strength of the electromagnetic interactions is at the order of weak interactions. 
Because the scalars have negative probability
these scalar fields cannot become physical scalar particles. 
Therefore, new physics is required at about $10^{14}$GeV.

Finally, it is worth to point out that the axial-vector currents and the charged vector currents 
of massive fermions and the top quark exist in all theories of EW interactions. 
Therefore, the three heavy scalars($m\sim10^{14}$GeV)
and the gauge fixings are model independent 
nonperturbative solutions of EW interactions.
 
%Astronomical data have provided strong evidences for an accelerated expanding universe at the present epoch.
%This acceleration is attributed to the domination($75\%$) of a component with negative pressure, dubbed dark energy.
%There are different approaches in studying dark energy. Scalar field can have negative $\rho+3P$($w<-{1\over3}$).
%Many models of scalar fields: Quintessence, K-essence, Phantom, and Quintom have been studied. 
%Negative kinetic energy is required for some of the scalar fields. Obviously, those scalar fields are not physical. 
%It is known that Higgs cannot be the candidate of dark energy.
%The scalar fields found in the EW theory have negative probability. 
%The Lagrangian of the scalar fields and interactions with 
%other fields can be derived from the EW theory. 
%The role played by the three heavy scalar fields in gravity and cosmology is an interesting and open question. 

\end{document}